\documentclass[amsmath,amssymb,preprint,showpacs]{revtex4}

\usepackage{graphicx}
\usepackage[usenames]{color}

\begin{document}

\title{Optical properties of the charge-density-wave polychalcogenide compounds $R_2$Te$_5$ ($R$=Nd, Sm and Gd)}
\author{F. Pfuner and L. Degiorgi} \affiliation{Laboratorium f\"ur
Festk\"orperphysik, ETH - Z\"urich, CH-8093 Z\"urich,
Switzerland}
\author{K.Y. Shin and I.R. Fisher}
\affiliation{Geballe Laboratory for Advanced Materials and
Department of Applied Physics, Stanford University, Stanford,
California 94305-4045, U.S.A.}

\date{\today}

\begin{abstract}
We investigate the rare-earth polychalcogenide $R_2$Te$_5$ ($R$=Nd, Sm and Gd)  charge-density-wave (CDW) compounds by optical methods. From the absorption spectrum we extract the excitation energy of the CDW gap and estimate the fraction of the Fermi surface which is gapped by the formation of the CDW condensate. In analogy to previous findings on the related $R$Te$_n$ (n=2 and 3) families, we establish the progressive closing of the CDW gap and the moderate enhancement of the metallic component upon chemically compressing the lattice.
\end{abstract}

\pacs{71.45.Lr,78.20.-e}


\maketitle

\section{Introduction}
The rare-earth polychalcogenides \cite{dimasi} $R$X$_n$ (where $R$ is the rare earth element, X  denotes S, Se and Te, and n=2, 2.5, 3) have recently attracted great interest due to their low dimensionality. Among the $R$X$_n$ families are members that variously host large commensurate distortions, ordered and disordered vacancy structures, and (small-amplitude) Fermi surface driven charge-density-wave (CDW). Furthermore, the discovery of a pressure-induced superconductivity state in CeTe$_2$ \cite{jung} competing with a CDW phase and an antiferromagnetic order makes the rare-earth tellurides an ideal system to investigate the consequences that the interplay or competition between those phases has on fermionic excitations at the Fermi energy. In a broader perspective, they may help providing a deeper understanding of how superconductivity might result from such an interplay, an issue of great interest in the solid state community. 

The CDW state has its origin in the well-known Peierls transition. Peierls first pointed out that a one-dimensional metal is unstable, when turning on the electron-phonon interaction, and undergoes a metal-insulator phase transition accompanied by a lattice distortion \cite{peierls}. The new modulation of the lattice induces a periodic potential, which will be screened by the itinerant charge carriers through the formation of the charge-density-wave condensate. Consequently, the opening of a (CDW) gap at the Fermi surface (FS) as well as the formation of a collective density wave state are the two most relevant fingerprints, characterizing a CDW broken symmetry ground state \cite{grunerbook}. The CDW gap acts as an order parameter of the phase transition and its determination is of relevance in order to get more insight about the impact of the transition on the electronic properties and Fermi surface, as well. 

Optical spectroscopic methods proved to be a powerful experimental tool and were widely used over the past few decades, in order to address the electrodynamic response in CDW materials \cite{grunerbook}. Recently, we have thoroughly studied the optical properties of the $R$Te$_2$ and $R$Te$_3$ series \cite{sacchettiprb,sacchettiprl,lavagniniprb}. They are closely related families, which are based on single ($R$Te$_2$) and double ($R$Te$_3$) Te-layers, separated by $R$Te block layers. We have discovered that the CDW gap is progressively reduced upon compressing the lattice in $R$Te$_3$ \cite{sacchettiprb,sacchettiprl}. This is accompanied by a release of additional charge carriers into the conducting channel, leading to a moderate enhancement of the Drude term (i.e., plasma frequency) in the optical spectra. Nonetheless, for both $n$=2 and $n$=3 series our optical findings establish that a large fraction of the Fermi surface is gapped by the CDW formation \cite{sacchettiprb,lavagniniprb}. In both families we have also identified a peculiar power-law behavior in the real part $\sigma_1(\omega)$ of the optical conductivity at frequencies higher than the CDW gap, compatible with predictions based on the Tomonaga-Luttinger liquid scenario. Besides implying the intrinsic one-dimensional nature of $R$Te$_2$ and $R$Te$_3$, the power-law behavior gives evidence for the important role played by correlation effects and Umklapp scattering processes in shaping the electronic properties of these materials, at least at high energies \cite{sacchettiprb,lavagniniprb}.

In comparison to $R$Te$_2$ and $R$Te$_3$, little is known about the $R_2$Te$_5$ family of compounds \cite{shin07}. Their orthorhombic crystal structure is intermediate between that of $R$Te$_2$ and $R$Te$_3$, comprising alternating single and double Te planes sandwiched between $R$Te corrugated block layers. The existence of this class of compound raises the question of whether separate modulation wave vectors might exist on the single and double Te planes, respectively, and if so how these wave vectors might interact or compete with each other. The origin of the CDW has been argued in terms of the electron instabilities through the calculation of the Lindhard susceptibility, based on the LMTO band structure \cite{shin07}. Recent transmission electron microscopy (TEM) investigations gave the first evidence for the CDW formation and established that the title compounds host a modulation wave vector, similar to that of the tri-telluride compounds. In addition, each $R_2$Te$_5$ compound exhibits at least one further set of superlattice peaks. Consideration of the electronic structure led to the conclusion that independent  wave vectors must be associated separately with sheets of the Fermi surface deriving from single and double Te layers, respectively \cite{shin07}. One could also speculate that even the CDW gap may be different from one layer to the other, eventually manifesting in a broad mid-infrared absorption. 

We present here our optical investigation on $R_2$Te$_5$, with $R$=Nd, Sm and Gd. The main goal is to shed light on their complete absorption spectrum in order to extract the relevant energy scales, as the CDW gap and the Drude plasma frequency. In a broader context, we also wish to establish, in parallel to structural considerations,  a comparison with the related rare-earth di- and tri-telluride families.

\section{Experiment and Results}

The single crystals of the investigated compounds were grown by slow cooling a binary melt. Details about their crystal growth and their characterization, particularly with respect to the CDW modulation vectors, can be found in Ref. \onlinecite{shin07}. As a consequence of a small exposed liquidus in the binary alloy phase diagram \cite{massalski}, crystals of $R_2$Te$_5$ grown by this technique often have a thin layer of $R$Te$_3$ on their outermost surface. This layer can be removed by cleaving using sticky tape, and x-ray experiments have revealed that the remaining $R_2$Te$_5$ crystal is phase pure. This issue is well illustrated in the inset of Fig. 1, which displays the spectra of Nd$_2$Te$_5$ as grown and after cleaving. The spectrum of the as-grown sample is indeed identical with the one of NdTe$_3$. One cleaving action is enough to remove the spurious $R$Te$_3$ phase at the surface. Repeated cleaving procedures do not change the optical response anymore. Similar to $R$Te$_3$, crystals of $R_2$Te$_5$ are readily oxidized, and care was taken to avoid prolonged exposure to air when preparing the specimens for our optical investigations. We could hinder aging effects by storing the samples in a clean atmosphere (glove box or inside our cryostat).

\begin{figure}[!tb]
\center
\includegraphics[width=8.5cm]{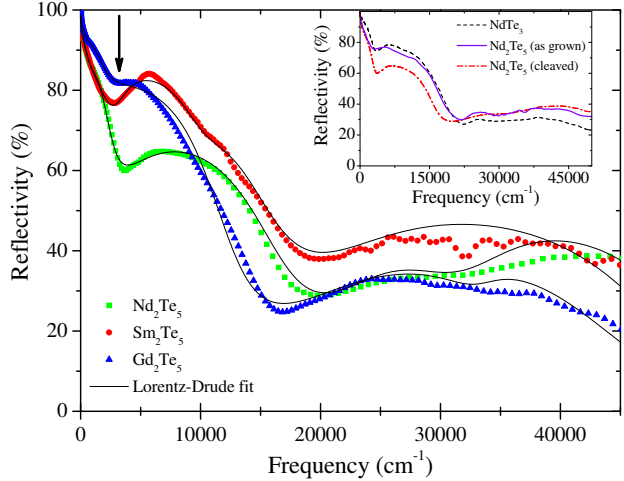}
\caption{(color online) Optical reflectivity of $R_2$Te$_5$ with $R$= Nd, Sm and Gd at 300 K. Thin lines are the Lorentz-Drude fit, as described in the main text. The arrow points out the depletion at about 4000 cm$^{-1}$. The inset compares the Nd$_2$Te$_5$ spectra, for the as-grown sample as well as for the cleaved one, with the optical response of the related NdTe$_3$ compound \cite{sacchettiprb} (see text).} \label{Refl.}
\end{figure}

We obtain the optical reflectivity $R(\omega)$ of all samples over a broad spectral range from the far-infrared up to the ultraviolet (50 - 50000 cm$^{-1}$  or 0.006 - 6 eV).
The complete optical response is achieved by combining three different spectrometers: for the far- and mid-infrared region we make use of the Bruker Fourier-Transform interferometer IFS 113, equipped with a Bolometer detector, as well as a Bruker (IFS 48) spectrometer, while for the visible and ultraviolet range we employ the Perkin Elmer Lambda 950 spectrometer. Further details pertaining to the experiment can be found elsewhere \cite{wooten,dressel}.

The main panel of Fig. 1 displays the $R(\omega)$ spectra for the three compounds at 300 K. The spectra turn out to be temperature independent between 10 and 300 K. Two features are clearly evident: the overall metallic behavior with a plasma edge onset at about 1.7x10$^4$ cm$^{-1}$, and a broad absorption, overlapped to the plasma edge and peaked around 5000 cm$^{-1}$. This latter absorption leads to a more or less pronounced depletion in $R(\omega)$ (arrow in Fig. 1), bearing striking similarities with previous data on $R$Te$_3$ \cite{sacchettiprb} and $R$Te$_2$ \cite{lavagniniprb}.

The real part $\sigma_1(\omega)$ of the complex optical conductivity is extracted from the Kramers-Kronig transformation of our reflectivity data \cite{wooten,dressel}. To this end, the data were appropriately extended to zero frequency ($\omega \rightarrow 0$) with the Hagen-Rubens extrapolation ($R(\omega)= 1-2\sqrt{\frac{\omega}{\sigma_{dc}}}$) and at high wavenumbers with a power-law extrapolation $\omega^{-s}$ (2$\leq  s \leq 4)$ \cite{wooten,dressel}. The $\sigma_{dc}$ values, inserted in the Hagen-Rubens formula, are in fair agreement with the dc transport results \cite{transportsmte}.
The resulting $\sigma_1 (\omega)$ for each compound is shown in Fig. 2, which emphasizes the features already pointed out in the reflectivity spectra (Fig. 1). For all three compounds, $\sigma_1 (\omega)$ is showing the so-called Drude peak in the low frequency range (with onset at frequencies $\omega \leq$ 1000 cm$^{-1}$), associated to excitations due to the free charge carriers, and a broad peak in the energy interval between 2000 and 8000 cm$^{-1}$.  Furthermore, $\sigma_1 (\omega)$ displays other features at high frequencies ($\omega \geq$ 10000 cm$^{-1}$), which can be tentatively ascribed to the onset of electronic interband transitions. Indeed, band structure calculations for the $R$Te$_n$ series \cite{shin07,laverock} suggest electronic excitations at energies above 10$^4$ cm$^{-1}$.

\begin{figure}[!tb]
\center
\includegraphics[width=8.5cm]{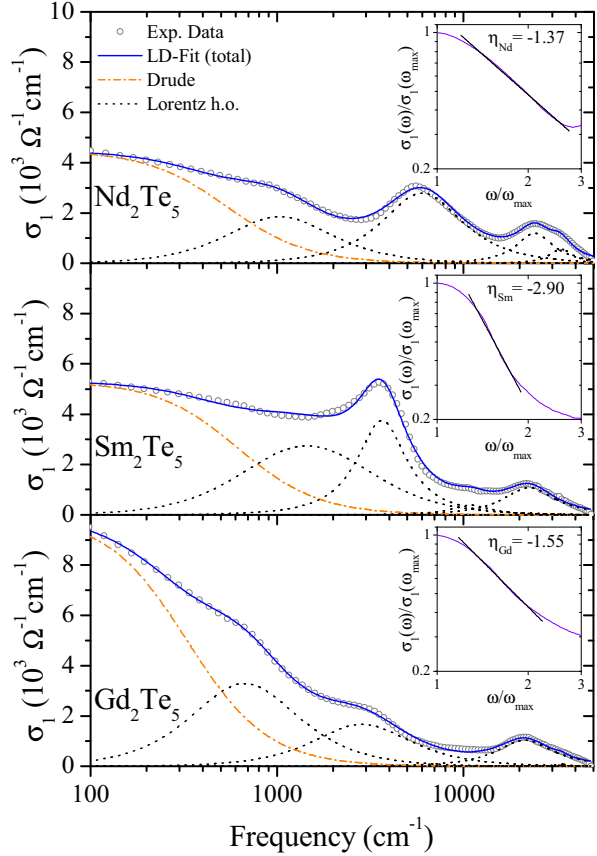}
\caption{(color online) Real part $\sigma_1(\omega)$ of the complex optical conductivity for the three title compounds (logarithmic energy scale). The total Lorentz-Drude (LD) fit as well as its components are shown for each compound. The insets display the normalized optical conductivity $\sigma_1(\omega)/\sigma_1(\omega_{max}$) versus $\omega/\omega_{max}$ (see text), emphasizing the power-law behavior in the spectral range above $\omega_{SP}$. The exponents $\eta$ are given in the figure as well as in Table I.} \label{sigma}
\end{figure}

For a detailed discussion of our results we apply the phenomenological  Lorentz-Drude approach. It is a common tool to analyse the optical response in condensed matter and consists first of all in reproducing the complex dielectric function by the following expression \cite{wooten,dressel}:
\begin{equation}
	\widetilde{\epsilon} (\omega)= \epsilon_1(\omega) + i \cdot \epsilon_2(\omega)= \epsilon_\infty - \frac{\omega^2_p}{\omega^2 + i \cdot \omega\cdot\gamma_D} + \sum_{n} \frac{S^2_n}{\omega^2 - \omega^2_n  - i\cdot\omega\cdot\gamma_n}.
\end{equation}
$\epsilon_\infty$ is the optical dielectric constant, $\omega_p$ the plasma frequency and $\gamma_D$ the scattering rate of the Drude term, whereas $S_n^2$, $\omega_n$ and $\gamma_n$ are the mode strength, the center frequency and the width of the n-th Lorentz harmonic oscillator (h.o.), respectively. From equation (1) we can then calculate all optical properties (e.g., the real part $\sigma_1 (\omega)= \omega \cdot \frac{\epsilon_2(\omega)}{4\cdot \pi}$ of the optical conductivity) as well as fully reproduce the measured $R(\omega)$ spectra. We systematically fit the optical spectra of the title compounds with four harmonic oscillators for the absorptions at finite frequencies, besides the Drude term for the metallic response. The fit components are shown in Fig. 2 for all investigated materials. The results of the fit in both $R(\omega)$ and $\sigma_1(\omega)$ are shown in Fig. 1 and 2, respectively, and testify a good agreement with the experiments. 

\section{Discussion}
The depletion at about 4000 cm$^{-1}$ in $R(\omega)$ (arrow in Fig. 1), overlapped to the plasma edge, and the corresponding mid-infrared absorption between 2000 and 8000 cm$^{-1}$ in $\sigma_1(\omega)$ are ascribed to the single particle peak (SP), due to the excitation across the CDW gap. This is in accordance with previous findings on related families; namely, $R$Te$_2$ and $R$Te$_3$ \cite{sacchettiprb,lavagniniprb}. ARPES results for Gd$_2$Te$_5$ give evidence for an energy scale of about 500 meV (4000 cm$^{-1}$) for the CDW gap \cite{shen}. Common to other rare-earth tellurides, the SP feature is rather broad, particularly as far as its low frequency side is concerned. This may suggest a possible distribution of gaps on FS, as also recognized in $R$Te$_2$ and $R$Te$_3$ by ARPES data \cite{shin,kang} and emphasized by our optical findings, as well \cite{sacchettiprb,lavagniniprb}. Similar to $R$Te$_2$ and $R$Te$_3$, the calculated FS for the unmodulated crystal structure of $R_2$Te$_5$ is imperfectly nested \cite{laverock2}, such that one can anticipate a range of gap sizes around the FS in the CDW state. There are perfectly nested regions with a large gap and non-perfectly nested ones with small gap or even with no gap at all. For the $R_2$Te$_5$ compounds, there is additionally the issue of the distinctly different lattice modulations, which may live on different parts of the crystal structure (single and double Te layer). The wave vectors for the two modulations might have different temperature dependence. The consequent wealth of gaps, observed in our optical view of the Brillouin zone, tends to spread out over a large energy interval, even merging into the high frequency tail of the metallic contribution. Obviously, we are not able optically to say anything firm about the gap distribution beyond this phenomenological guess, since we are essentially integrating over the entire FS. 

\begin{table}[h]
	\centering
		\begin{tabular}{|c|c|c|c|c|}
		\hline
 $R_2$Te$_5$ \phantom{\Huge{1}}					   & 	$\omega_{SP}$  	& 	$\omega_p$  	&		$\Phi$		&		$\eta$ \\
\hline
\hline
Nd$_2$Te$_5$  \phantom{\Huge{1}}     &		5308			&		12050		&		0.18		&		-1.4 \\
\hline
Sm$_2$Te$_5$ \phantom{\Huge{1}}    &		2640				&		14000		&		0.15		&		-2.9 \\
\hline
Gd$_2$Te$_5$ \phantom{\Huge{1}} &		2115			& 	14050		&		0.23		&		-1.6 \\
		\hline
		\end{tabular}
		\caption{Rare-earth $R$ dependence of the single particle peak $\omega_{SP}$, the plasma frequency $\omega_p$ (all entries in cm$^{-1}$), the ratio $\Phi$ and power-law exponent $\eta$.} \label{Tab}
\end{table}

From a pure phenomenological point of view, the gap distribution or, more straightforward, the broad SP is fitted with two Lorentz h.o.'s in eq. (1) \cite{comment}, while the remaining two h.o.'s at high frequencies (i.e., $\omega\ge$ 10$^4$ cm$^{-1}$) mimic the absorptions due to the interband transitions (Fig. 2). In analogy to our previous analysis \cite{sacchettiprb,lavagniniprb}, it seems again convenient to introduce the so-called weighted energy scale $\omega_{SP}$:
\begin{equation}
\omega_{SP}=\frac{\sum_{j=1}^2 \omega_j S_j^2}{\sum_{j=1}^2
S_j^2},
\end{equation}
where the sum extends over the first two h.o.'s of the mid infrared absorption. $\omega_{SP}$ represents the center of mass of the SP excitation. The values for $\omega_{SP}$ along with the plasma frequency $\omega_p$ are summarized in Table I. The optical findings on $R_2$Te$_5$ confirm the overall trend in both optical parameters, already encountered in related rare-earth telluride families. Upon compressing the lattice there is a diminishing impact of the CDW phase, which is reflected in the decrease of $\omega_{SP}$ as well as in the moderate enhancement of $\omega_p$. Progressively closing the gap also means that additional free charge carriers are released into the conduction channel, thus increasing the weight of the Drude term. Figure 3a displays the energy scale $\omega_{SP}$ as a function of the in-plane lattice constant $a$ for the rare-earth telluride $R$Te$_n$ ($n$= 2, 2.5 and 3) series. It is quite appreciable that all these compounds are characterized by a common reduction of the CDW gap upon compressing the lattice. Moreover, it is worth recalling that the progressive suppression of the CDW gap upon compressing the lattice has been recently discovered in the optical properties of CeTe$_3$ \cite{sacchettiprl} and LaTe$_2$ \cite{lavagninipre} under externally applied pressure.

\begin{figure}[!tb]
\center
\includegraphics[width=8.5cm]{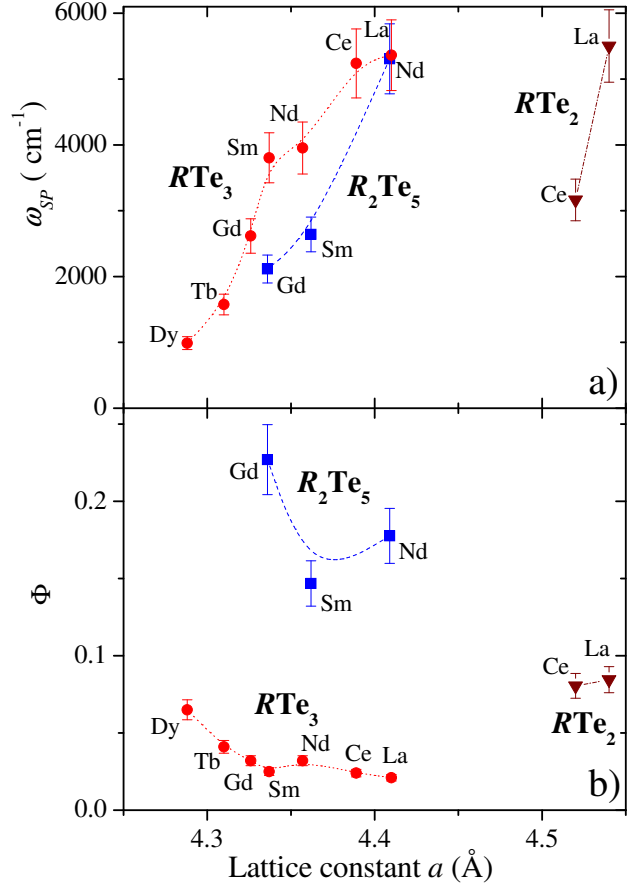}
\caption{(color online) a) The single particle peak frequency $\omega_{SP}$ (eq. 2) and b) the ratio $\Phi$ (eq. 3) are shown as a function of the in-plane lattice constant $a$ for the $R_2$Te$_5$, $R$Te$_3$ \cite{sacchettiprb} and $R$Te$_2$ \cite{lavagniniprb} series.} \label{wSP}
\end{figure}

Sum rule arguments allow us to estimate the fraction of FS, affected by the formation of the CDW condensate. Following our previous work on NbSe$_3$ \cite{perucchi}, $R$Te$_2$ \cite{sacchettiprb} and $R$Te$_3$\cite{lavagniniprb}, we can define the ratio:
\begin{equation}
\Phi=\omega_p^2/(\omega_p^2+\sum_{j=1}^2 S_j^2) 
\end{equation}
between the Drude weight in the CDW state and the total spectral weight of the hypothetical normal state. This latter quantity is achieved by assuming that above $T_{CDW}$ the weight of the single particle peak (i.e., $\sum_{j=1}^2 S_j^2$) merges together with the Drude weight. Equation (3) provides a measure of how much of the FS survives in the CDW state. The values of $\Phi$ for $R_2$Te$_5$ are displayed in Table I, while Fig. 3b visualizes the comparison of the ratio $\Phi$ between the three families of compounds. The overall increase in $\Phi$ with chemical pressure for $R_2$Te$_5$ (i.e., of about 20\% from the Nd to the Gd compound) follows a similar trend to that which is observed for $R$Te$_3$ \cite{sacchettiprb}, even though the fraction of the ungapped FS in the CDW state appears to be larger in $R_2$Te$_5$ than in other polychalcogenides (Fig. 3b). The non-monotonic behavior of $\Phi$ in $R_2$Te$_5$ could be ascribed to the variation in lattice modulations between the three compounds, in contrast to $R$Te$_3$ for which the lattice modulation is the same. Sm$_2$Te$_5$ is in this regard quite peculiar among the $R_2$Te$_5$ materials, since it is characterized by two independent CDWs, commensurate along the in-plane $c$-axis and incommensurate along the in-plane $a$-axis. These CDWs are pertinent to the Te double and single layers, respectively \cite{shin07}. It is believed furthermore that the two CDWs as well as the two types of Te layers interact more in the Sm compound than in the other two.

The tri- as well as the di-telluride compounds have shown some evidence for a power-law behavior ($\sigma_1(\omega)\sim \omega^{\eta}$) on the high frequency tail of the SP feature \cite{sacchettiprb,lavagniniprb}. While the power-law extends in a rather limited spectral range (particularly in the case of the di-tellurides \cite{lavagniniprb}), electronic correlation effects were advocated as a possible origin for this peculiar behavior in $\sigma_1(\omega)$. Similarly to the linear chain organic Bechgaard salts \cite{vescoliscience}, we have envisaged that the direct electron-electron interaction, as source for Umklapp scattering, leads to a novel non-Fermi liquid quantum state, based on the Tomonoga-Luttinger liquid concepts. The typical fingerprints within this theoretical framework would indeed include the power-law behavior of $\sigma_1(\omega)$ at energies above the CDW gap \cite{giamarchi}. If applicable to these families of compounds, the Tomonaga-Luttinger liquid scenario would also imply a (hidden) one dimensional character of the electronic properties, at least at the high frequencies corresponding to the optical measurements. 

The insets of Fig. 2 display a normalized view of the optical conductivity (i.e., $\sigma_1(\omega)/\sigma_1(\omega_{max})$ versus $\omega/\omega_{max}$, where $\omega_{max}$ is the peak frequency of the SP feature) in the spectral range pertinent for the power-law behavior. The data are suggestive of power-law behavior over a very small frequency range (much
less than a decade), but the presence of additional interband transitions at higher frequencies precludes definitive statements. The estimates of the exponent $\eta$ are summarized in Table I, with an error of about $\pm$ 0.1. The values of $\eta$ range from -1.4 and -1.6 for Nd$_2$Te$_5$ and Gd$_2$Te$_5$ respectively (compatible with a
weakly interacting limit and quite different to -2, as expected for the high frequency decay of $\sigma_1(\omega)$ due to a single Lorentzian), to -3 for Sm$_2$Te$_5$ (more appropriate for the case of a band insulator where the lattice is assumed to be rigid and only Umklapp scattering off the single-period lattice potential is possible \cite{giamarchi,vescoli}). We emphasize that caution should be placed on these values of $\eta$ given the small frequency range over which they were extracted.

\section{Conclusions}
We provided here thorough optical investigations of the electrodynamic response of three representative members of the $R_2$Te$_5$ family of compounds, which share several common features with previous findings on related polychalcogenides. The CDW gap decreases upon compressing the lattice, thus generalizing concepts already developed for the $R$Te$_2$ and $R$Te$_3$ series. The presence of single and double Te-layers in the crystal structure of  $R_2$Te$_5$ \cite{shin07} considerably affects the nesting properties of the Fermi surface as well as the impact of the CDW condensate on the electronic properties of these materials. In this context one might eventually explain the subtle differences among the title compounds as far as the FS gapping and the effect of low-dimensional correlations are concerned. It remains to be seen how one can reconcile the different power-law behavior for the three title compounds within an overall CDW scenario.

\begin{acknowledgments}
The authors wish to thank J. M\"uller for technical help, T. M\"uller for her contribution in the first stage of the data collection, and A. Sacchetti and M. Lavagnini for fruitful discussions. This work has been supported
by the Swiss National Foundation for the Scientific Research
within the NCCR MaNEP pool. This work is also supported by the
Department of Energy, Office of Basic Energy Sciences under
contract DE-AC02-76SF00515.
\end{acknowledgments}


\begin{thebibliography}{99}
\bibitem{comment} In contrast to $R$Te$_3$ \cite{sacchettiprb} and $R$Te$_2$ \cite{lavagniniprb}, two Lorentz h.o.'s are enough in $R_2$Te$_5$ to phenomenologically reproduce the SP mid-infrared feature defining the CDW gap. The h.o.'s mainly account for the spectral weight encountered in the SP absorption. The distribution of CDW gaps, evinced from ARPES \cite{shen,shin}, might also be addressed by an almost continuous distribution of Lorentz h.o.'s. Such an approach would go however beyond the scope of the Lorentz-Drude fit, because of the large amount of fit parameters thus introduced.
\end{thebibliography}
\end{document}